\documentclass[aps,twocolumn,showpacs,preprintnumbers,amsmath,amssymb]{revtex4}

\usepackage{graphicx}
\usepackage{dcolumn}
\usepackage{bm}

\begin{document}

\title{Skew band structure and anomalous conductivity of PdCrO$_2$}

\author{Yu.B.Kudasov}
\email{yu_kudasov@yahoo.com}
\affiliation{Sarov Physics and Technology Institute, National Research Nuclear University "MEPhI", 
Dukhov str. 6, Sarov, 607188, Russia}
\affiliation{Russian Federal Nuclear Center - VNIIEF, Mira str. 37, Sarov, 607188, Russia}

\date{\today}
\begin{abstract}
A model of magnetic interaction of CrO$_2$ and Pd hexagonal layers in PdCrO$_2$ is proposed. Since
Cr-O-Pd bridges do not provide an interlayer magnetic coupling in case of a 120$^0$ magnetic
ordering, a direct exchange interaction between magnetic chromium ions and 
conductive palladium layers is assumed.  
It is shown that this interaction leads to a novel state of itinerant electrons (skew bands). It is
characterized by abnormally high conductivity at low temperatures due to strong suppression of umklapp 
electron-phonon scattering.
\end{abstract}


\maketitle

Hexagonal layered ABO$_2$ compounds with the delafossite type structure demonstrate a variety of exotic 
phenomena: frustration and complex magnetic phase diagram with noncollinear
magnetic structures \cite{CuFeO}, multiferroic behavior \cite{CuFeAlO}, anomalous transport properties 
\cite{Mackenzie}, and etc.
During the last decade an intense interest was attracted
to unusual electronic transport in PdCoO$_2$, PtCoO$_2$, and PdCrO$_2$ \cite{Eyert,Mackenzie}.
Their room-temperature conductivity reaches
the highest value among oxide metals and approaches that of elementary metals such as aluminum,
copper, and silver \cite{Mackenzie}. A mean free path in PdCoO$_2$ is 700~{\AA} at the room temperature 
and
increases up to about 20~$\mu$m (or 10$^5$ lattice periods) at low temperatures \cite{Hicks}. This implies 
a novel mechanism of electronic transport because a characteristic distance between lattice imperfections 
is definitely a few orders of magnitude shorter than this value. 
The long-lived momentum of electrons causes an emergent hydrodynamic regime of electron transport in these 
compounds \cite{hydro,moll}.

A structure of the metallic delafossites consists of 2D hexagonal layers of palladium or platinum, which 
provide electron transport, separated by isolating CoO$_2$ or CrO$_2$ spacers \cite{Lechermann, Ong}. 
Chromium ions form a complex magnetic order in PdCrO$_2$ below $T_N\approx$38~K: a 120$^0$ magnetic 
structure appears in each CrO$_2$ layer with staggered chirality in neighboring layers \cite{Takatsu,Le}. 
Totally there are 18 magnetic sublattices. The intralayer magnetic structure
can be described by a pseudodipole model which arises from superexchange interactions in the presence of 
spin-orbit coupling \cite{Le}.
The CrO$_2$ layers are bound with each other by dumbbells O-Pd-O. However, in case of the 120$^0$ 
intralayer magnetic ordering the interlayer magnetic interactions between the sublattices through the 
dumbbells cancel each other \cite{Takatsu}. 
That is why, a weak ring interaction \cite{Park} was proposed
as a source of the interlayer coupling.

The resistivity demonstrates a sharp drop when PdCrO$_2$ undergoes the transition to the 3D magnetic order 
at $T_N$ \cite{Takatsu2},
that is, the magnetism promotes the high-conductivity state in this compound. An anomalous behavior of 
magnetothermopower indicates that a magnetic short-range order with a correlation length much greater than
the lattice period persists well above $T_N$ \cite{mtp}. Single-crystal
neutron diffraction measurements also show development of two-dimensional magnetic correlations above 
$T_N$
\cite{Billington}. 

Another intriguing problem is an unconventional anomalous Hall effect (UAHE) observed
in PdCrO$_2$ \cite{Takatsu3}. The 120$^0$ magnetic structure with staggered chirality
gives the zero total chirality. Thus, PdCrO$_2$ provides a rare example of UAHE occurring at zero total 
chirality \cite{Takatsu}.  

The Fermi surface in the metallic delafossites was thoroughly investigated 
\cite{Hicks,Ok,Hicks2,Billington}. It was found to be nearly two-dimensional in all the compounds. The 
Fermi surface in PdCoO$_2$ has a rounded hexagonal cross-section corresponding to a half-filled conduction 
band \cite{Hicks}. The non-collinear magnetic ordering causes its $\sqrt{3}\times\sqrt{3}$ reconstruction 
in PdCrO$_2$. As a result, a sheet of the reconstructed Fermi surface, which is close to the magnetic 
Brillouin zone boundaries ($\gamma$ orbit), appears\cite{Ok,Hicks2}. There are also pockets ($\alpha$ 
orbits) in the corners of the magnetic Brillouin zone.
   
\begin{figure*}
	\includegraphics[width=0.15\textwidth]{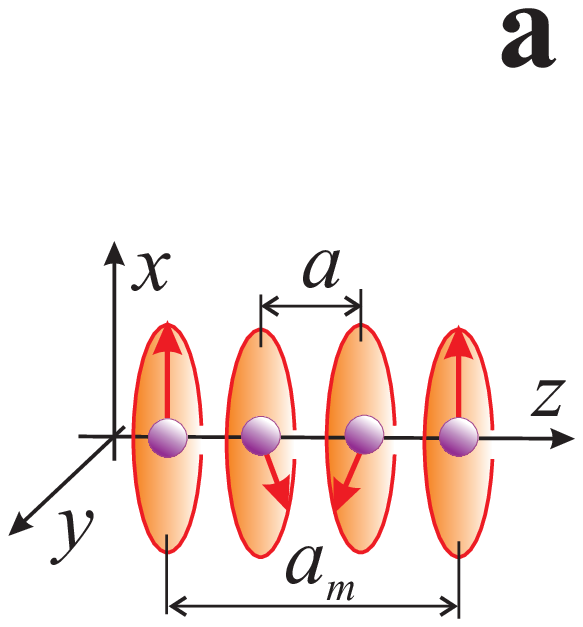}
	\hfill
	\includegraphics[width=0.35\textwidth]{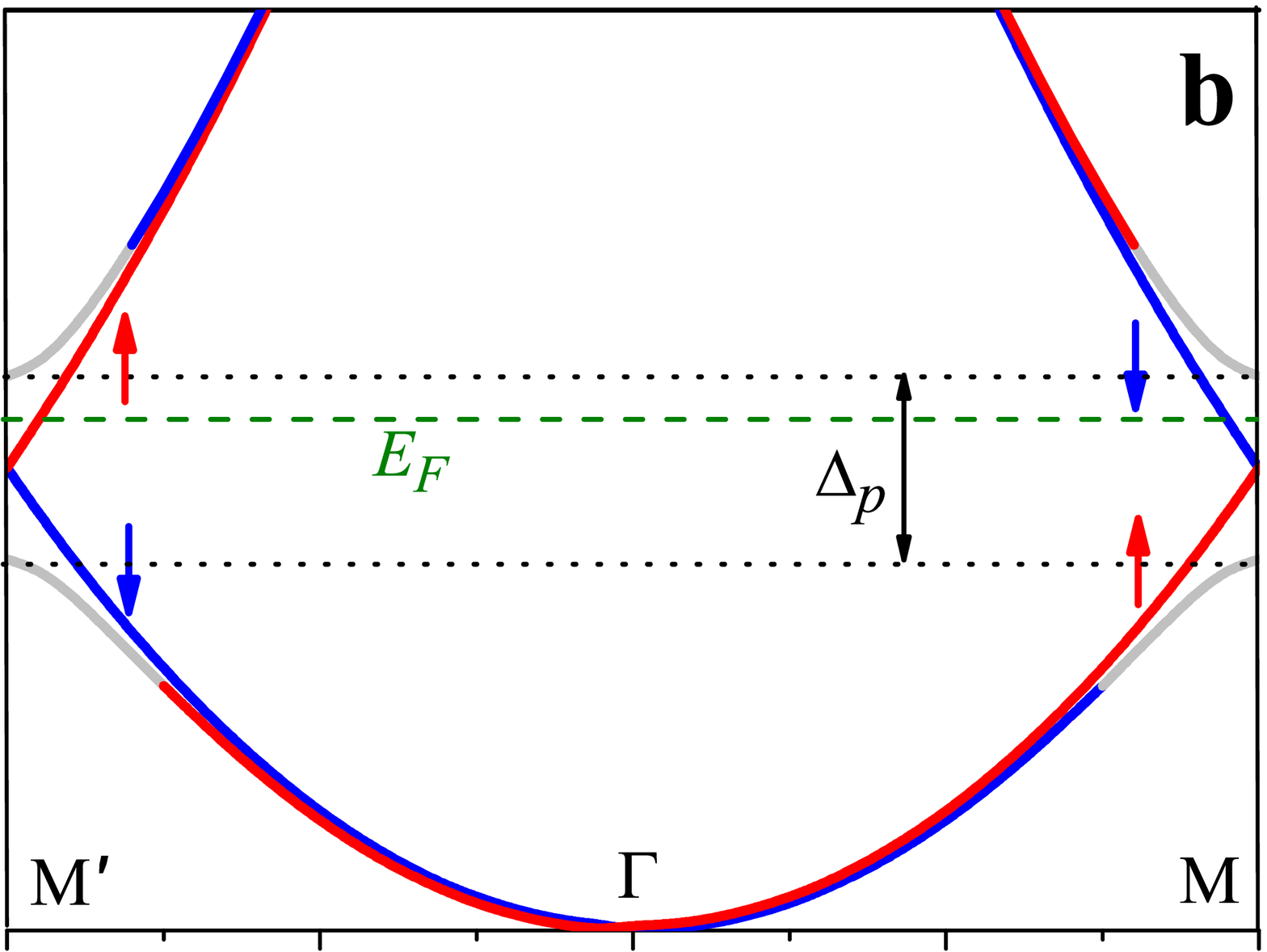}
	\hfill
	\includegraphics[width=0.35\textwidth]{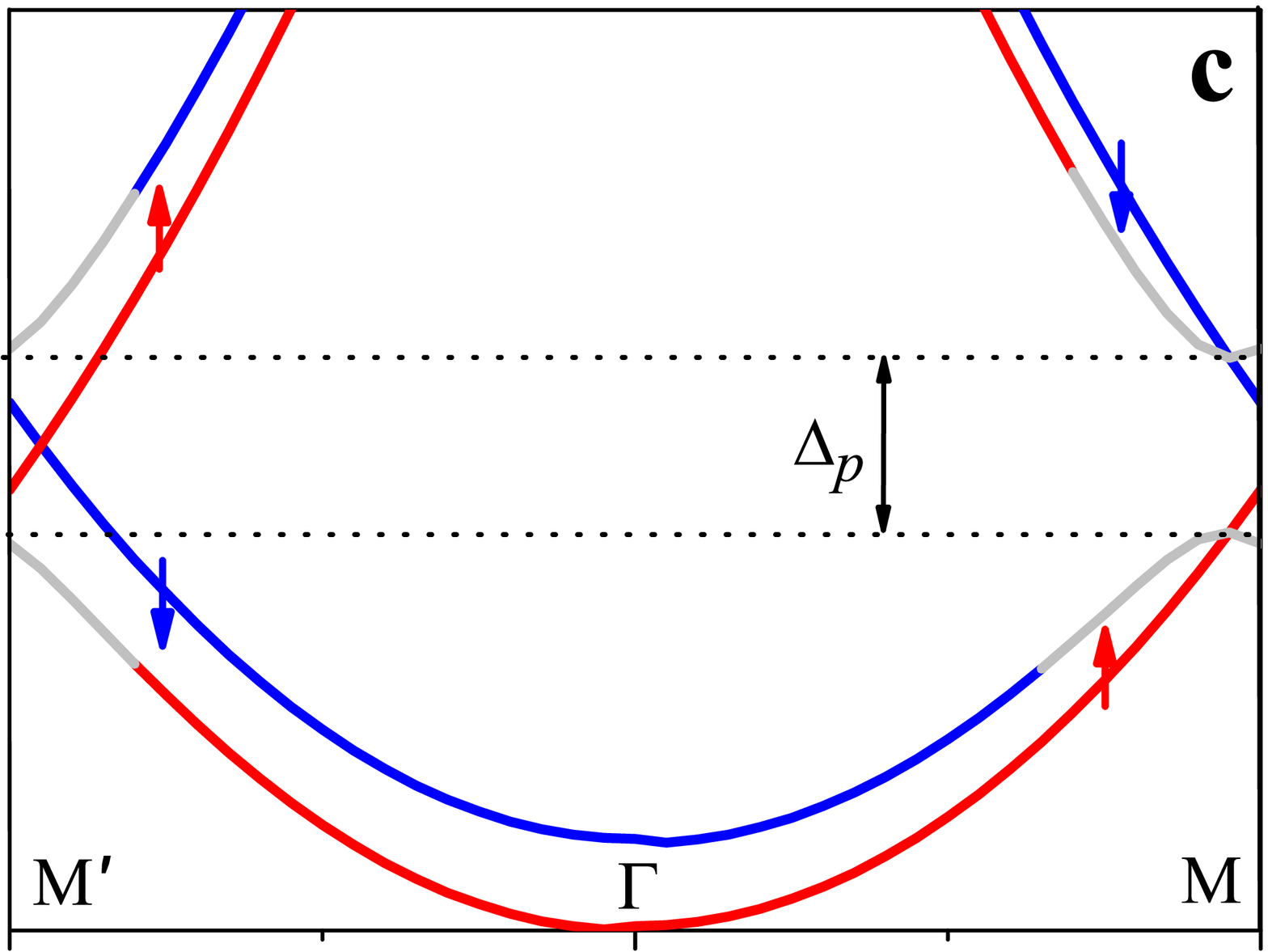}
	\caption{\label{f1} (color online) 1D structure under the spiral magnetic field: (a) a schematic view,
	 (b) the dispersion curves for $h_z=0$, and (c) the same for $h_z=h_0/2$. The arrows and color denote
	 the up- and down-spin states, strongly mixed spin states (see Eq.~(\ref{ansatz})) are marked with the
	 gray lines.}
\end{figure*}

Let us start with a simple 1D model to illustrate an effect of periodical screw-spiral magnetic field on 
itinerant electrons. The Hamiltonian of the system has the following form 
\begin{equation}
\left[ \hat{H}_0 + \mathbf{h}\left( \mathbf{r} \right) \hat{\bm{\sigma}} \right] \Psi = \mathcal{E} \Psi
\label{Ham}
\end{equation}
where $\hat{H}_0 = - \Delta/2 + V(\mathbf{r})$ is the nonmagnetic part of the Hamiltonian, i.e. the sum of
the kinetic energy and a periodic crystal potential $ V(\mathbf{r})$, $\mathbf{h}\left( \mathbf{r} 
\right)$ is the magnetic field, $ \hat{\bm{\sigma}}$ are the Pauli matrices, and $\Psi$ is the 
two-component spinor. We assume the 1D structure is oriented along the $z$-axis ($\mathbf{r} \equiv z$, 
$\Delta \equiv \partial^2 \big/ \partial z^2 $). The magnetic
field lies in the $xy$-plane and has the form of a spiral with a period $a_m$ which is a multiple of
the crystal period $a$ ($a_m>a$), as shown in Fig.~\ref{f1}a: $h_x(z)=h_0 \cos(K z)$ and $h_y(z)=h_0 
\sin(K z)$ where $h_0$ is a constant, $K=2\pi/a_m$.

Fourier coefficients for the magnetic potential are obtained by integration over the magnetic unit
cell     
\begin{eqnarray}
\hat{U}_{\mathbf{K}} = \frac{1}{V} \int \exp(-i \mathbf{K} \mathbf{r}) \mathbf{h}\left( \mathbf{r} \right) 
\hat{\bm{\sigma}} \mathrm{d}\mathbf{r} \label{UK} =\frac{h_0}{2} \left(\hat{\sigma}_x -i \hat{\sigma}_y 
\right). \label{UKb}
\end{eqnarray}
Since $\hat{\bm{\sigma}}$ are Hermitian operators, $\hat{U}_{\mathbf{-K}} = \hat{U}_{\mathbf{K}}^\dagger$,
$\mathbf{K}$ is the reciprocal-lattice vector.
Considering the magnetic potential as a weak periodic perturbation one can apply the simple
ansatz
\begin{eqnarray}
| \Psi_{\mathbf{k}i} \rangle =  C_{\mathbf{k} i \sigma} | \mathbf{k}, \sigma \rangle +
C_{\mathbf{k} - \mathbf{K} i \bar{\sigma}} | \mathbf{k} - \mathbf{K}, \bar{\sigma} \rangle
\label{ansatz} 
\end{eqnarray}
where $ | \mathbf{k}, \sigma \rangle$ is the  unperturbed state, $C_{\mathbf{k}i \sigma}$ are the complex 
coefficients, $\bar{\sigma}$ is the value of spin projection opposite to $\sigma$, and $i=1,2$. Then the 
dispersion curves  
can be determined by a traditional manner \cite{Ashcroft}
\begin{eqnarray}
\left( \hat{\bm{\epsilon}}_\mathbf{k} -  \varepsilon_\mathbf{k}^0 \hat{\mathbf{I}}  \right)\left( 
\hat{\bm{\epsilon}}_\mathbf{k} - \varepsilon_{\mathbf{k}-\mathbf{K}}^0 \hat{\mathbf{I}} \right) =
\hat{U}_{\mathbf{K}} \hat{U}_{\mathbf{K}}^\dagger
\label{det}
\end{eqnarray}
where $\hat{\bm{\epsilon}}_\mathbf{k}$ is the unitary matrix with eigenvalues $\varepsilon_{\mathbf{k}1}$
and  $\varepsilon_{\mathbf{k}2}$, $\hat{\mathbf{I}}$ is the unit matrix,
$\varepsilon_\mathbf{k}^0$ is the electrons energy of
the unperturbed (nonmagnetic) system ($\hat{H}_0$).   

The dispersion curves within the magnetic Brillouin zone for $V(\mathbf{r})=0$ are shown Fig.~\ref{f1}b.
A pair of non-degenerate bands with opposite spins appears in the pseudo-gap $\Delta_p= 2 h_0$ which is 
shown by the horizontal dot lines in Fig.1b. They deserve a closer look.
These branches coincide with $\varepsilon_\mathbf{k}^0$, and also $\varepsilon_{\mathbf{k}\sigma} = 
\varepsilon_{-\mathbf{k}-\sigma}$. Such relations usually take place in systems with a spin-orbit 
interaction.
In fact, the bands are non-symmetric with respect to $\Gamma$ point (skew bands) but form a mutually 
symmetrical pair. This is clearly seen if we add a uniform magnetic field along the $z$ axis by entering 
an additional term $\hat{\sigma}_z h_z$ in both expressions in brackets in Eq.~(\ref{det}) (see 
Fig.~\ref{f1}c). 

The 1D skew band structure demonstrates unusual features. It is easy to see that if the Fermi level $E_F$ 
lies in the pseudo-gap, e.g. as shown in Fig.1b, an elastic backward scattering without spin-flip is 
forbidden
and there exists a persistent spin current.

\begin{figure}
	\includegraphics[width=0.18\textwidth]{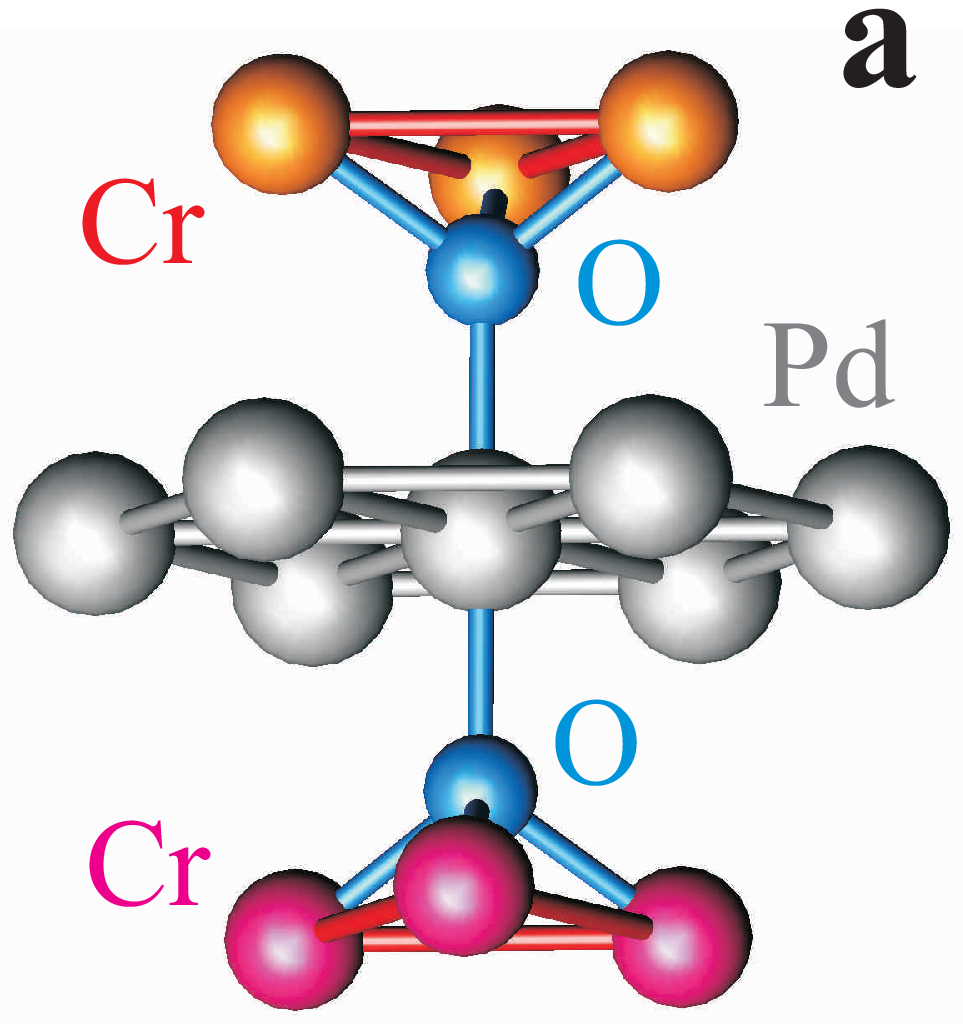}
	\hfill
	\includegraphics[width=0.23\textwidth]{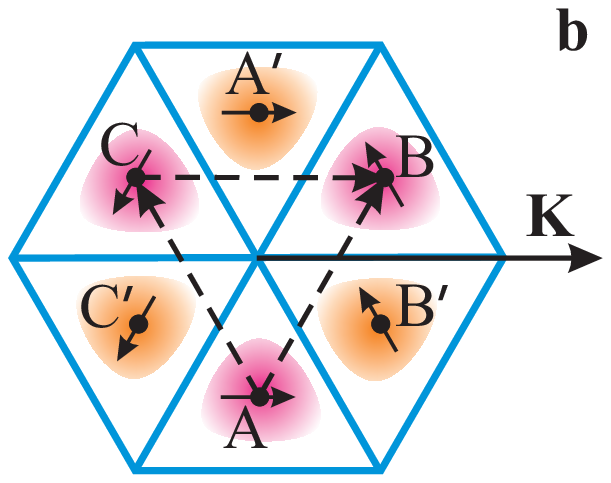}
	\caption{\label{f2} (color online) (a) The schematic view of Pd hexagonal layer with adjacent CrO$_2$ 
layers, (b) the top view of magnetic unit cell: the position labels correspond to Cr superlattices (see 
the text), directions of magnetization of chromium ions are shown by the small arrows.}
\end{figure}

A strong effect of magnetic ordering in a chromium layer on itinerant electrons in an adjacent palladium 
layer is provided by Cr-O-Pd
bridges as shown in Fig.~\ref{f2}a. However, the three chromium ions of the top or bottom layer belong to 
three different magnetic sublattices and, in case of
the 120$^0$ intralayer magnetic order,
exactly neutralize each other's action. That is why, much weaker interactions are responsible for the 
coupling of magnetic layer and itinerant electrons. 

We assume that a direct exchange interaction of chromium ions and itinerant electrons induces an effective 
field in the palladium layer. It is described by the same Hamiltonian Eq.~(\ref{Ham}) where $\mathbf{r}$ 
is the 2D vector in the plane of the palladium layer.
Firstly, let us consider an effect of single chromium layer, say the bottom layer in Fig.~\ref{f2}a. Then 
the effective field takes the form
\begin{equation}
 \mathbf{h}\left(\mathbf{r}\right)= \sum_{i=A,B,C}{ f\left(\mathbf{r}-\mathbf{r}_i \right)  \mathbf{m}_i 
\hat{\bm{\sigma}}}
\label{efffield}
\end{equation}
where the indices $A,B,C$ denote the chromium magnetic sublattices, $f\left(\mathbf{r}\right)$ is the 
integrand in the exchange integral,  
$\mathbf{r}_i$ is the projection of position of the chromium nuclei onto the palladium plane, 
$\mathbf{m}_i$ is the magnetization of the $i$-th ion. The positions of the ions
and function $f\left(\mathbf{r}\right)$ are schematically shown in  Fig.~\ref{f2}b. For the sake of 
simplicity we assume that $\mathbf{m}_i$ lie in the $xy$ plane \cite{xy}.

The vectors $\mathbf{r}_{AB}=\mathbf{r}_{B}-\mathbf{r}_{A}$, 
$\mathbf{r}_{CA}=\mathbf{r}_{C}-\mathbf{r}_{A}$, and $\mathbf{r}_{CB}=\mathbf{r}_{C}-\mathbf{r}_{B}$ are 
the translation vectors of
the non-magnetic lattice. At the same time, they correspond to one third of the translation vectors of the 
three-sublattice magnetic structure. Then the integration in Eq.~(\ref{UK}) gives the following expression
\begin{eqnarray}
\hat{U}_{\mathbf{K}} = \sum_{i=A,B,C}{ F \alpha_i \mathbf{m}_i \hat{\bm{\sigma}}} \label{UK2} 
\end{eqnarray}
where $F= \int \exp\left( -i \mathbf{K} \mathbf{r}\right)  f\left(\mathbf{r}-\mathbf{r}_A\right) 
\mathrm{d}\mathbf{r} / V$ and $\alpha_i$ are coefficients which have the following allowed values: $1$ and 
$\exp\left(\pm 2 \pi/3 \right) $. In case of the vector $\mathbf{K}$ oriented as shown in
Fig.~\ref{f2}b  we obtain $\alpha_A = 1$, $\alpha_B = \exp\left(- 2 \pi/3 \right) $, and $\alpha_C = 
\exp\left(2 \pi/3 \right) $.

The magnetizations of chromium sublattices corresponding to the 120$^0$ magnetic order 
are given by $\mathbf{m}_A = m_0\left( \mathbf{i} \cos(\phi) + \mathbf{j} \sin(\phi)\right) $,
$\mathbf{m}_B = m_0\left( \mathbf{i} \cos(\phi+2\pi \chi /3) + \mathbf{j} \sin(\phi+2\pi \chi /3)\right) 
$, and $\mathbf{m}_C = m_0\left( \mathbf{i} \cos(\phi-2\pi \chi /3) + \mathbf{j} \sin(\phi-2\pi \chi 
/3)\right) $. Here $\mathbf{i}$ and $\mathbf{j}$ are the unit basis vectors in the $xy$ plane, $\chi= \pm 
1$ is the chirality, and $\phi$ is the initial angle \cite{Takatsu}. The Fourier coefficients for the 
magnetic potential are determined by substituting these expressions in Eq.~(\ref{UK2}) :
\begin{eqnarray}
\hat{U}_{\mathbf{K}} = \frac{3}{2} e^{i \chi \phi} F m_0 \left(\hat{\sigma}_x -i  \chi \hat{\sigma}_y 
\right). 
\label{UK1}
\end{eqnarray}
This equation matches Eq.~(\ref{UKb}) up to an unessential factor. Then dispersion curves along the 
M'--$\Gamma$--M line are the same as those of the 1D model (Fig.~\ref{f1}b).
Applying the ansatz (\ref{ansatz}) one obtains the pseudo-gap width $\Delta_p = 3 F m_0$ regardless of the 
chirality and initial angle.
A change of the chirality sign merely alternates
the spin of the in-gap states.
The substitution
$\mathbf{K} \rightarrow \mathbf{-K}$ leads to the same effect. 

To make the model more realistic for describing the electronic structure of PdCrO$_2$ let us consider a 
pair of magnetic layers adjacent to the palladium
layer (Fig.~\ref{f2}a). It is easy to show by direct calculation that, if the chirality of the top and 
bottom magnetic layers are opposite, one obtains
an ordinary band gap at a boundary of the magnetic Brillouin zone. In case of the same chirality the band 
structure
turns out to be similar to that for a single magnetic layer. The only difference is that the pseudo-gap 
becomes anisotropic and depends on the difference $\phi_t - \phi_b$ , where the indices $t$ and $b$ stand 
for the top and bottom layers. For example, if $\phi_t = \phi_b = 0$
the vector $\mathbf{K}$ lies along the symmetry line for the magnetization of the top and bottom layers as 
shown in Fig.~\ref{f2}b, the pseudo-gap is $\Delta_p = 12 F m_0$. For the other orientations it remains 
$\Delta_p = 3 F m_0$ .

\begin{figure}
	\includegraphics[width=0.25 \textwidth]{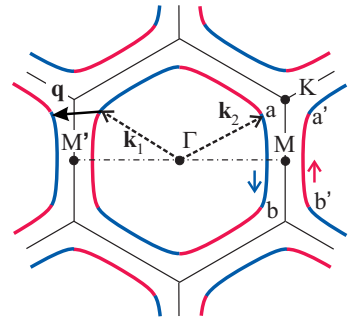}
	\caption{\label{f3} (color online) The Fermi surface of skew band structure. The spin states are 
marked by color and small arrows. The initial ($\mathbf{k}_1$), final ($\mathbf{k}_2$), and phonon 
($\mathbf{q}$) wave vectors are shown by the dash and solid arrows.}
\end{figure}

Let us assume the same chirality of the top and bottom magnetic layers.
In fact, as was mentioned above, neutron diffraction in PdCrO$_2$ revealed staggered chirality in chromium 
layers \cite{Takatsu}. This inconsistency is discussed below. The reconstructed Fermi surface
in PdCrO$_2$ is close to boundary of the magnetic Brillouin zone ($\gamma$ orbit). This means that the 
Fermi level lies in the pseudo-gap, e.g. as shown in Fig.~\ref{f1}b by the dash line. The corresponding 
shape of the Fermi surface is depicted in Fig.~\ref{f3}. It consists of alternating areas with opposite 
spins that shown by colors and arrows. In the vicinity of the $\Gamma$--K line the ansatz (\ref{ansatz}) 
is incorrect and should be extended up to three terms including two reciprocal lattice vectors. That is 
why, the spin states intermix at
the intersection of the $\Gamma$--K line and Fermi surface.
  
To consider transport properties of the 2D skew bands (Fig.~\ref{f3}) we assume as usual  that the 
temperature dependence of conductivity at low temperatures is determined by electron-phonon interaction 
\cite{Ashcroft,Ziman}. We do not  dwell on a normal scattering because it is similar to that in ordinary 
metals. It is well-known \cite{Ziman} that an umklapp scattering determines a momentum relaxation of 
electron-phonon system as a whole and, therefore, electrical conductivity. 
The umklapp processes give rise to the conductivity term of the following form \cite{vanVucht}: $\rho_U 
\propto T^n \exp\left(- \varepsilon_{\mathbf{q}m}/T\right)$ where $T$ is the temperature, 
$\varepsilon_{\mathbf{q}m}$ is the phonon energy corresponding to the minimal wave vector, $n$ is the 
constant depending on Fermi surface shape and phonon spectrum.

At low temperatures, phonons with small wave vectors are involved in
scattering processes only. Then the umklapp scattering between the arcs ab and a'b' in Fig.~\ref{f3}b 
should give the main
contribution to $\rho_U$ \cite{Ziman}. However, a general form of the  electron-phonon Hamiltonian of the 
first and second orders in atomic displacements conserves an electron spin \cite{Grimvall}. That is why, 
transitions between the arc facing each other are forbidden.
The umklapp scattering with small phonon wave vectors $\mathbf{q}$ is allowed only in the vicinity of K 
points as shown in Fig.~\ref{f3}. It involves small areas in the corners of the Fermi surface $\Delta k 
\sim T/u$ where $u$ is the sound velocity (acoustic phonons are considered only). This reduces $\rho_U$ in 
the skew bands by a multiplier $\sim T/T_D$. Another factors lowering the umklapp resistivity should be 
mentioned. (i) The angle between the initial ($\mathbf{k}_1$) and final ($\mathbf{k}_2$) wave vectors of 
electron in the umklapp process is about $\pm 2 \pi /3$, that is backward umklapp scattering is forbidden. 
(ii) The scattering vector $\mathbf{k}_2 - \mathbf{k}_1$
is almost collinear to the phonon wave vector $\mathbf{q}$
that strongly suppresses scattering by low-energy transverse phonons \cite{Grimvall,Kaveh}.  

When the umklapp electron-phonon scattering is weak the low-temperatures resistivity is determined by 
$\rho_U$ due to the phonon drag effect \cite{Kaveh}:  $\rho = \gamma \rho_U$
where $\gamma$ is a factor of the order of unity. Thus, the total electron-phonon resistivity in PdCrO$_2$ 
occurs to be very low.  

As was mentioned above the skew bands appear only in case of
the same chirality in the adjacent magnetic layers. However, the neutron single crystal and synchrotron 
X-ray powder diffraction experiments clearly indicate the staggered chirality in PdCrO$_2$ 
\cite{Takatsu,Le}.
On the other hand, the magnetic Bragg peaks width shows that along the $z$ axis the magnetic structure is 
less correlated than in
the plane (the correlation length is about 97~\AA) \cite{Billington}. Therefore there exist palladium 
layers at the domain boundaries which have neighboring magnetic layers of the same chirality. They provide 
the high conductivity of the compound. This assumption can be verified, for instance, by means of 
impedance spectroscopy. 

A two-dimensional short-range magnetic order is observed in PdCrO$_2$ above $T_N$ \cite{Billington}. It 
persists up to rather high temperatures \cite{mtp}. That is why, the proposed mechanism of anomalous 
conductivity is also applicable to the paramagnetic phase of this substance. 

Although PdCoO$_2$ and PtCoO$_2$ are nonmagnetic an anomalous Hall effect attributed to Stoner's surface 
magnetic layers \cite{Harada} as well as extremely high magnetoresistance \cite{Takatsu4} indicate a role 
of magnetic correlations. It should be mentioned that although Co$^{3+}$ ions are nominally non-magnetic 
in an octahedral environment they
can provide a strong indirect exchange interaction (e.g. in Ca$_3$Co$_2$O$_6$ \cite{kudasov}). Thus, the 
model proposed can be extended to paramagnetic compounds with short-range magnetic order.

In conclusion, the skew band structure in a palladium layer of PdCrO$_2$ is induced by a pair of 
neighboring magnetic layers with 120$^0$ order of the same chirality. The resistivity of the Pd layer 
turns out to be very low because the umklapp electron-phonon scattering is drastically suppressed.
A backward scattering without spin-flip is forbidden. A detailed discussion of the skew band structure 
will be presented elsewhere.

I gratefully acknowledge fruitful discussions with A. N. Vasil'ev, A. A. Fraerman, and V. V. Platonov.  

\bibliography{kudasov}

\begin{thebibliography}{27}
\bibitem[{xy()}]{xy}
\bibinfo{note}{In fact, the spin plane in PdCrO$_2$ is vertical, i.e. it
  contains the $z$ axis \cite{Takatsu}. However, the simplification used does
  not affect the final result.}
\end{thebibliography}

\end{document}